\documentstyle[12pt,epsf]{article}

\begin{document}

% useful macros
\newcommand {\bea}{\begin{eqnarray}}
\newcommand {\eea}{\end{eqnarray}}
\newcommand {\be}{\begin{equation}}
\newcommand {\ee}{\end{equation}}

\def\IR{{\hbox{{\rm I}\kern-.2em\hbox{\rm R}}}}
\def\IH{{\hbox{{\rm I}\kern-.2em\hbox{\rm H}}}}
\def\IC{{\ \hbox{{\rm I}\kern-.6em\hbox{\bf C}}}}
\def\IZ{{\hbox{{\rm Z}\kern-.4em\hbox{\rm Z}}}}

\begin{titlepage}
\rightline{EFI-2000-5}

\rightline{hep-th/0002166}

\vskip 3cm
\centerline{\Large{\bf Kaluza-Klein Black Holes in String Theory}}
\vskip 2cm
\centerline{Finn Larsen}
\vskip 12pt
\centerline{\sl the Enrico Fermi Institute}
\centerline{\sl University of Chicago}
\centerline{\sl 5640 S. Ellis Ave., Chicago, IL 60637, USA}
\centerline{\texttt{flarsen@theory.uchicago.edu}} 
\vskip 2cm

\begin{abstract}
Non-supersymmetric black holes carrying both electric and magnetic charge 
with respect to a single Kaluza-Klein gauge field have much in common with 
supersymmetric black holes. Angular momentum conservation and other general 
physics principles underlies some of their basic features. Kaluza-Klein black 
holes are interpreted in string theory as bound states of D6-branes and D0-branes. 
The microscopic theory reproduces the full nonlinear mass formula of the 
extremal black holes. 
\end{abstract}
\end{titlepage}

\newpage

\section{Introduction}
Black holes are interpreted microscopically in string theory as bound 
states of explicitly specified constituents. It is therefore an important 
theoretical challenge to identify examples where the quantum bound state 
problem can be analyzed. The canonical example is the bound state of D1- and 
D5-branes where the microscopic theory is known in 
great detail~\cite{Strominger:1996sh}. In this and many other cases the asymptotic 
degeneracy of states has been determined to agree with the Bekenstein-Hawking formula 
for the black hole entropy. In all cases where such an agreement has been 
established with precision the near-horizon geometry of the black hole 
contains an $AdS_{3}$ factor; and this feature underlies the 
agreement~\cite{Strominger:1998eq,Balasubramanian:1998ee}. It remains an open 
problem to similarly understand black holes with less symmetric near-horizon 
geometries; especially black holes with no supersymmetry. The present work reports on 
progress in this direction, for a specific case.

Consider the original Kaluza-Klein theory in four dimensions, obtained by 
compactification of five-dimensional pure gravity on a circle. The field 
content of this theory is a $U(1)$ gauge field, a scalar field, and gravity.
Stationary black holes solutions to the theory are parametrized by 
their electric (Q) and magnetic (P) gauge charges, as well as their mass 
(M) and angular momentum (J). There is a simple embedding of the 
system into string theory, as follows. First, add six compact dimensions, 
{\it e.g.} a Calabi-Yau three-fold, or a six-torus. Next, interpret the original 
Kaluza-Klein direction as the M-theory circle so that the electric Kaluza-Klein 
charge is identified as D0-brane charge, and the magnetic Kaluza-Klein charge 
similarly becomes the charge of a D6-brane, fully wrapped 
around the six inert dimensions. {\it The black hole is therefore 
interpreted at weak coupling as a bound state of D0-branes and 
D6-branes}. 

In string theory it is the norm to consider black holes with many independent 
charges excited simultaneously. The electric and magnetic charges are 
thus generalized to vectors. It has been the experience that black holes with 
non-orthogonal charge vectors pose special difficulties: the microscopic description 
is less constrained~\cite{Dijkgraaf:1997it,Larsen:1999dh}, and the corresponding 
classical solutions are much more 
complicated~\cite{Cvetic:1996bj,Cvetic:1995kv,Bertolini:1999je}. An elementary 
property of the Kaluza-Klein dyon considered here is that its electric and magnetic 
charge vectors -- being numbers -- have nonvanishing inner-product, 
${\vec P}\cdot{\vec Q}=PQ\neq 0$. The Kaluza-Klein black hole is therefore a simple 
setting where these problems can be analyzed. In fact, to the present author, this 
was the original motivation for considering the problem. The study of Kaluza-Klein 
black holes is further motivated by the ``no-frill'' character of the system, by 
fundamental string theory interest in the D0/D6 bound state
~\cite{Taylor:1997ay,Lifschytz:1996bh,Brandhuber:1997tt,Dhar:1998ip,branco,Itzhaki:1998ka}, 
and as a means to study black holes~\cite{Khuri:1996xq,Ortin:1996bz,Sheinblatt:1998nt}.

The black hole metric can be constructed explicitly for arbitrary 
$(M,J,Q,P)$~\cite{Rasheed:1995zv,Larsen:1999pp,Matos:2000ai}. This will not be 
repeated here. Instead the emphasis will be on the qualitative properties of the 
system. The presentation is based on the article~\cite{Larsen:1999pp}, 
except for a new result(given in section~\ref{sec:micro}), the microscopic 
interpretation of the mass formula for the bound state .

\section{Properties of Extreme Black Holes}
\label{bh}
For a given value of the conserved charges $Q,P,J$ there is a lowest 
possible value of the mass $M$ consistent with regularity. The three 
parameter family of solutions saturating this bound are the {\it extremal}
black holes. The physical interpretation is that extremal black holes 
in some sense are in their ground states. This presentation considers only 
the extremal case.

\subsection{Basic Parameters}

\paragraph{Black hole mass:} 
First, assume that the rotation is limited according to $G_{4}J<PQ$. In 
this ``slow rotation'' case the mass formula is:
\be
2G_{4}M = (Q^{2/3}+ P^{2/3})^{3/2}~.
\label{eq:extmass} 
\ee
This mass formula has been known for some time, for $J=0$~\cite{Gibbons:1986ac};
the striking point emphasized here is the {\it independence} of angular 
momentum. In other words, the system can carry angular momentum at no 
cost in energy. Other features of the extremal geometry {\it do} depend on the 
angular momentum, as expected. 

Next, assume that $G_{4}J>PQ$. In this 
``fast rotation'' case the mass formula is more complicated, the solution of a 
quartic equation. The mass formula now depends on the angular momentum as well as 
the charges. It satisfies:
\be
2G_{4}M > (Q^{2/3}+ P^{2/3})^{3/2}~.
\label{eq:lextmass}
\ee
The two branches of extremal black holes are joined by a two-parameter family 
of black holes satisfying $G_{4}J=PQ$. The geometry degenerates in this 
critical limit; for example, the black hole entropy approaches zero.

\paragraph{Black hole entropy} is also sensitive to the
boundary at $G_{4}J=PQ$. Indeed, for slow rotation $G_{4}J<PQ$:
\be
S = 2\pi\sqrt{ {P^{2}Q^{2}\over G_{4}^{2}} - J^{2}}~,
\label{eq:extent}
\ee
while for fast rotation $G_{4}J>PQ$:
\be
S = 2\pi\sqrt{ J^{2}-{P^{2}Q^{2}\over G_{4}^{2}}}~.
\label{eq:exent}
\ee
The only change is thus the overall sign under the square root. The 
extremal Kerr-Newman black hole and its four-parameter generalization 
canonically considered in string 
theory~\cite{Kallosh:1996uy,Horowitz:1996ac,Cvetic:1996kv}, coincide for special 
choices of parameters with the {\it fast} rotation case 
(\ref{eq:exent}). Here the main 
interest is the {\it slow} rotation case (\ref{eq:extent}). 

\paragraph{The black hole temperature} vanishes in the 
extremal limit for {\it all} values of the angular momentum. 
This conforms with general expectations for extremal black holes.

\subsection{Comparison with Supersymmetric Black Holes}
The fast rotating Kaluza-Klein black holes are very similar to extremal 
Kerr-Newman black holes in four dimensions. A more surprising analogy
is between the slowly rotating Kaluza-Klein black holes and the rotating 
BPS black holes in five dimensions, interpreted as excitations of 
D1/D5-brane bound states~\cite{Breckenridge:1996is}. The striking similarities
include: 

\paragraph{1)} In the D1/D5-case the energy of a supersymmetric ground state 
is related to its momentum by supersymmetry, ensuring that the black hole mass 
is independent of the angular momentum. The D1/D5-system therefore also has 
the property that it can carry angular momentum at no cost in energy.

\paragraph{2)} The supersymmetric ground states of the D1/D5 system are generally 
charged under the R-charge of the supersymmetry algebra, which is 
identified with the spacetime angular momentum~\cite{Breckenridge:1996is}. 
The projection on to a given value of the R-charge restricts the 
available phase space, and so decreases the entropy; it vanishes when 
the angular momentum is so large that all states are forced to have 
identical projection of the angular momentum. The black hole entropy has the 
form:  
\be
S = 2\pi\sqrt{{1\over 4}J_{3} - J^{2}}~,
\label{eq:d1d5ent}
\ee
where $J_{3}$ is the unique cubic invariant of $E_{6(6)}$.
This should be compared with the entropy (\ref{eq:extent}), or more 
generally the U-duality invariant expression:
\be
S = 2\pi\sqrt{{1\over 4}J_{4} - J^{2}}~,
\label{eq:stent}
\ee
where $J_{4}$ is the unique quartic invariant of $E_{7(7)}$.
The similarity suggests that the Kaluza-Klein black holes are described by 
a supersymmetric conformal field theory with a structure similar to the one 
familiar from the D1/D5-system. Specifically, the angular momentum should be 
identified with an R-charge in such a description.

\paragraph{3)} For slowly rotating Kaluza-Klein black holes
the angular velocity of the horizon 
{\it vanishes} $\Omega_{H}=0$ . The physical interpretation is 
that the angular momentum is carried by the field surrounding the black hole, 
rather than by its interior. The unfamiliar combination of angular momentum, 
but no angular velocity, occurs also for the rotating BPS black holes in five 
dimensions. For fast rotation, the horizon velocity remains finite in the 
extremal limit $\Omega_{H}\neq 0$, as for the Kerr black hole. 

\paragraph{4)} Outside the horizon of a rotating black hole there is 
an {\it ergosphere}. This is a region where observers cannot remain 
at rest relative to the asymptotic geometry, because the drag of the geometry 
force them to rotate along with the black hole. Such observers are nevertheless 
free to escape to infinity. An important consequence of the ergosphere is that it
allows the black hole to shed rotational energy classically by 
{\it superradiance}. This effect renders {\it e.g.} the standard extremal 
Kerr black hole in four dimensions unstable. The D1/D5-system corresponds 
to rotating black holes in five dimensions and for these, remarkably, the 
ergosphere disappears in the extremal limit. This saves the stability of the 
system required by supersymmetry. Interestingly, the ergosphere of the 
Kaluza-Klein black hole {\it also} disappears in the extremal limit, for 
slow rotation. On the other hand, for fast rotation there {\it is} an ergosphere;
so the black hole decays classically, even though it is extremal. In a sense, 
the mass (\ref{eq:lextmass}) on the large rotation branch is too large, and the 
black hole seeks to reach the lower bound (\ref{eq:extmass}) which apparently 
is more stable. 

\vspace{1.cm}
At this point it may appear that slowly rotating Kaluza-Klein black holes 
are precisely analogous to the D1/D5-system. That is far from the truth. 
The D1/D5 system is supersymmetric; indeed, it is the only case familiar to me 
where rotation is consistent with the BPS 
condition~\cite{Chamseddine:1997pi,Gauntlett:1998kc}. Many of 
the remarkable properties discussed above follow from this fact. It is thus 
significant to emphasize that Kaluza-Klein black holes are {\it not} 
supersymmetric.

To see this, embed the simple Kaluza-Klein theory in a theory with at least 
N=2 supersymmetry. The supersymmetry algebra then implies:
\be
2G_{4}M \geq \sqrt{Q^{2}+P^{2}}~,
\label{eq:msusy}
\ee
with the inequality saturated if and only if the black hole preserves a part 
of the supersymmetry. The mass formulae (\ref{eq:extmass}-\ref{eq:lextmass})
satisfy this condition; however,they never saturate it, when both 
electric and magnetic charges are present. Kaluza-Klein black holes 
are therefore {\it not} supersymmetric.

Without supersymmetry, the question of stability should be considered 
seriously. The energy of two widely separated fragments, each carrying 
either the electric or the magnetic charge is:
\be
2G_{4}M \geq Q+P~.
\ee
This inequality is also satisfied by the mass formulae 
(\ref{eq:extmass}-\ref{eq:lextmass}); so spontaneous fragmentation of the 
black hole into two parts is consistent with energy conservation. 

However, in the present system it is important to consider also angular 
momentum conservation. The electric and magnetic fragments are charged 
with respect to the {\it same} $U(1)$ gauge field; so the total angular 
momentum of the final state satisfies Dirac's bound:
\be
J \geq {PQ\over G_{4}}~.
\label{eq:dirac}
\ee
The lower bound coincides precisely with the one classifying Kaluza-Klein 
black holes as having slow or fast rotation. It is interesting that the 
evident qualitative distinction between slow and fast rotation is 
related to the Dirac bound on the angular momentum: the geometry ``knows'' 
about the Dirac bound. Concretely, the bound implies that angular momentum 
conservation {\it forbids} decay of the slowly rotating black holes 
into two widely separated electric and magnetic fragments; but the fast 
rotating ones {\it do} decay in this way. 

It is possible that the slowly rotating black hole instead decays into two 
widely separated {\it dyons}, with charge assignments $(Q_{1},P_{1})$ and 
$(Q_{2},P_{2})$, respectively. The Dirac bound (\ref{eq:dirac}) on 
the angular momentum of the fragments is then replaced by:
\be
J\geq {|P_{1}Q_{2}-P_{2}Q_{1}|}/G_{4}~.
\ee
For example two identical dyons can have vanishing angular momentum. 
There are still large classes of black holes that have no possible 
decays; for example non-rotating black holes with mutually prime quantized 
charges. In fact, standard stability arguments, using supersymmetry, 
are similarly subject to conditions on the quantum numbers of the state.
The possible decay into two dyons is therefore consistent with the analogy 
between BPS states and the slowly rotating branch. 

These results do not imply that the slowly rotating black holes are 
absolutely stable. For example, the angular momentum could be carried 
away by one of the decay products. An example that realizes this 
possibility is the Callan-Rubakov effect~\cite{Callan:1982ah,Rubakov:1981rg}; 
here a charged spin-1/2 fermion interacts with a monopole, but the combined system 
nevertheless supports a spin-0 mode. An alternative decay channel involves a third 
particle carrying spin, but arbitrarily low energy; {\it e.g.} a graviton. Despite 
the existence of allowed decay channels, it is evident that the slowly rotating 
extremal black holes exhibit a remarkable degree of stability; in particular, it 
is suggestive that the most obvious decay channel is forbidden. It would be 
interesting to make a stronger and more precise statement on this issue.

\subsection{Quantization Rules}
Up to this point, the electric and magnetic charges have been 
arbitrary parameters. After embedding into quantum theory they 
are quantized:
\bea
Q &=& 2G_{4}M_{0}~n_{Q}~, \label{eq:Qdef}\\
P &=& 2G_{4}M_{6}~n_{P}~, \label{eq:Pdef}
\eea
where $n_{Q}$ and $n_{P}$ are {\it integral}.  In the D0/D6 
interpretation discussed in the introduction:
\bea
M_{0}&=& {1\over l_{s}g_{s}}~,\label{eq:M0def}\\
M_{6}&=& {V_{6}\over (2\pi)^{6}l_{s}^{7}g_{s}}~,\label{eq:M6def}
\eea
where the string units are defined so $l_{s}=\sqrt{\alpha^{\prime}}$ and 
$V_{6}$ is the volume of the six compact dimensions wrapped by the 
$D6$-brane; as always $G_{4}={1\over 8}(2\pi)^{6}l^{8}_{s}g^{2}_{s}$. 
This gives the relation $8G_{4}M_{0}M_{6}=1$ (which in fact is expected
from general principles). Thus:
\be
{2PQ\over G_{4}}= n_{Q}n_{P}~.
\label{eq:dquant}
\ee
As a check on normalizations note that, after this quantization condition 
is taken into account, the lower bound in (\ref{eq:dirac}) 
quantizes the angular momentum as a half-integer. A related point is that the 
entropy (\ref{eq:extent}-\ref{eq:exent}) simplifies. After the quantization 
condition is taken into account it is expressed in terms of pure numbers, 
{\it i.e.} the moduli cancel out. This is promising for a connection to 
microscopic ideas.

\section{The Microscopic Description}
\label{sec:micro}
The analogy with BPS black holes suggests that it is possible to describe 
Kaluza-Klein black holes precisely in the underlying string theory. As discussed 
in the introduction, the microscopic interpretation of the Kaluza-Klein black 
hole is a bound state of $k=n_{Q}$ D0-branes and $N=n_{P}$ D6-branes. 
The theory on the D6-branes is a field theory in 6+1 dimensions. The field content 
is the same as maximally supersymmetric Yang-Mills theory with $SU(N)$ gauge 
group. D0-branes are described in this theory as excitations with third Chern-class 
equal to the number of D0-branes, and vanishing first and second Chern-classes.
Assuming that the compact dimensions span a six-torus, it is simple to 
construct examples of this kind using time-independent field strengths of the 
form:
\be
F_{12} = f\mu_1~;~~~~
F_{34} = f\mu_2~;~~~~
F_{56} = f\mu_3~,
\label{eq:Fdefs}
\ee
where the $SU(N)$ matrices $\mu_i$ satisfy:
\be
\mu_i^2 = I~;~~~~
\mu_1 \mu_2 \mu_3 = I~;~~~~
{\rm Tr} \mu_i = 0~;~~~~
{\rm Tr} \mu_i \mu_j = 0~.
\label{eq:mudefs}
\ee
The first and second Chern-classes vanish because the trace of $F$,
and also of $F\wedge F$, vanish along all cycles. The third Chern-class -- 
and so the number of D0-branes -- is:
\be
k = {1\over 6(2\pi)^3}\int {\rm Tr} F\wedge F\wedge F = {1\over (2\pi)^3}
N V_6 f^3~.
\ee
It is convenient to use (\ref{eq:Qdef}-\ref{eq:M6def}) and rewrite 
this relation as $(2\pi)^3 l_s^6 f^3 = Q/P$.

The D6-brane wraps a small compact manifold, so it is legitimate to ignore 
higher derivatives in the action. The interactions of the theory are 
therefore given by the Born-Infeld Lagrangean. For static configurations the 
corresponding mass functional is:
\be
M = T_6 \int {\rm Tr}\sqrt{{\rm det}\left( 1 +  2\pi l_s^2 F \right)}~,
\ee
where $T_{6}$ is the tension of the $D6$-brane, {\it i.e.} its mass 
density. For the explicit configurations given above the mass becomes:
\be
M = T_6 V_6 N \left(1 + (2\pi)^2 l^4_s f^2\right)^{3/2}=
\left( P^{2/3}+Q^{2/3}\right)^{3/2}/(2G_4)~.
\ee
This is precisely the mass formula (\ref{eq:extmass}). A similar computation was 
presented in~\cite{Taylor:1997ay}, for charges $N=k=4$ and moduli chosen such that 
$P=Q$. The generalization given here shows that a full functional dependence is 
reproduced by the microscopic considerations.

\section{Discussion}
The D0/D6-interpretation of the system is valid at weak coupling, {\it i.e.}
when the ambient spacetime is mildly curved. In contrast, the black 
hole description applies at strong coupling. The agreement between the mass formulae
obtained in the two mutually exclusive regimes therefore suggests a 
duality. Indeed, although Kaluza-Klein theory does {\it not} admit an $SL(2,\IZ)$ 
duality group, it preserves a $\IZ_{2}$ subgroup interchanging electric and magnetic 
charges, as well as weak and strong coupling. The mass formula is therefore not 
necessarily invariant under extrapolation from weak to strong coupling; but the 
two regimes are related by a discrete symmetry.

It is simple to construct explicit microscopic configurations of the form 
(\ref{eq:Fdefs}-\ref{eq:mudefs}); indeed, there are numerous ways to do so.
Moreover, if the {\it ansatz} is relaxed, there are additional possibilities. 
The system therefore has considerable microscopic degeneracy which is presumably 
related to the black hole entropy. However, a precise confirmation of 
this idea has not yet been achieved.

Another open problem concerns the world-volume interpretation of the angular 
momentum. As discussed after (\ref{eq:stent}), the black hole 
entropy formula suggests a relation to the R-charge of a superconformal 
algebra; or at least some $U(1)$ current in a 2D conformal field theory. 
Unfortunately it seems difficult to identify a specific current with the required 
properties. 

{\bf Acknowledgments:}
I thank V. Balasubramanian, A. Goldhaber, P. Kraus, J. Harvey, J. Maldacena, 
E. Martinec, and A. Peet for discussions; and the Niels Bohr 
Institute for hospitality during the preparation of the manuscript.
This work was supported by DOE grant DE-FG02-90ER-40560 and by a Robert R. McCormick 
Fellowship.

\end{document}